\documentclass[cameraready]{Interspeech}

\title{FastTurn: Unifying Acoustic and Streaming Semantic Cues for Low-Latency and Robust Turn Detection}

\author[affiliation={1}, equalcontribution]{Chengyou}{Wang}
\author[affiliation={1}, equalcontribution]{Hongfei}{Xue}
\author[affiliation={1}]{Mingchen}{Shao}
\author[affiliation={1}]{Chunjiang}{He}
\author[affiliation={1}]{Jingbin}{Hu}
\author[affiliation={1}]{Shuiyuan}{Wang}
\author[affiliation={2}]{Bo}{Wu}
\author[affiliation={2}]{Yuyu}{Ji}
\author[affiliation={2}]{Jimeng}{Zheng}
\author[affiliation={2}]{Ruofei}{Chen}
\author[affiliation={1}, correspondingauthor]{Lei}{Xie}

\address{
    $^1$ Audio, Speech and Language Processing Group (ASLP@NPU) \\
    $^2$ Shengwang 
}

\email{asd6404112a@mail.nwpu.edu.cn, lxie@nwpu.edu.cn}


\usepackage{comment}
\usepackage{multirow}
\usepackage[table]{xcolor}
\usepackage{balance}
\definecolor{lightgreen}{RGB}{230,245,230}

\begin{document}

\maketitle

\begin{abstract}

Recent advances in AudioLLMs have enabled spoken dialogue systems to move beyond turn-based interaction toward real-time full-duplex communication, where the agent must decide when to speak, yield, or interrupt while the user is still talking. Existing full-duplex approaches either rely on voice activity cues, which lack semantic understanding, or on ASR-based modules, which introduce latency and degrade under overlapping speech and noise. Moreover, available datasets rarely capture realistic interaction dynamics, limiting evaluation and deployment. To mitigate the problem, we propose \textbf{FastTurn}, a unified framework for low-latency and robust turn detection. To advance latency while maintaining performance, FastTurn combines streaming CTC decoding with acoustic features, enabling early decisions from partial observations while preserving semantic cues. We also release a test set based on real human dialogue, capturing authentic turn transitions, overlapping speech, backchannels, pauses, pitch variation, and environmental noise. Experiments show FastTurn achieves higher decision accuracy with lower interruption latency than representative baselines and remains robust under challenging acoustic conditions, demonstrating its effectiveness for practical full-duplex dialogue systems.

\end{abstract}

\section{Introduction}

In recent years, rapid advances in AudioLLMs~\cite{huang2024audiogpt,chu2024qwen2,yu2024salmonn,zeng2024glm,xie2024mini,geng2025osum} have enabled spoken dialogue systems to move beyond traditional turn-based interaction toward more natural real-time communication. In highly interactive scenarios, this evolution leads to full-duplex interaction, where the system must process speech perception, partial semantic understanding, and response planning concurrently while the user is still speaking. Unlike turn-based settings~\cite{lu2025duplexmamba}, a full-duplex system is required to make online decisions about when to continue speaking, when to yield the floor, and when to insert or interrupt~\cite{liu2020towards,leviathan2018google,defossez2024moshi}. These decisions involve a delicate latency–accuracy trade-off: reacting too late increases overlap and errors, while reacting too early risks truncating semantics and degrading coherence, especially under noisy and overlapped observations. Although large language models excel at reasoning and generation with complete textual inputs, integrating them into low-latency full-duplex dialogue systems remains challenging.

In practical deployments, existing full-duplex spoken dialogue systems~\cite{leviathan2018google,defossez2024moshi} often rely on turn detection to provide a controllable interface between speech processing and response generation. Current turn detection approaches can be broadly categorized into two groups. The first group relies on voice activity detection (VAD) and infers interruption timing from acoustic energy or activity patterns~\cite{wang2024freeze, chen2025fireredchat, fu2025vita}. These methods are lightweight and fast, but primarily capture speech presence rather than communicative intent, making them prone to false triggers from backchannels, hesitations, or background noise. The second group introduces explicit turn prediction modules using learned models. Representative examples include Smart Turn, TEN Turn Detection, and Easy Turn~\cite{DBLP:journals/corr/abs-2509-23938}. The second approach enhances conversational intent detection by leveraging learned models and text-based cues, making it more adaptable to complex dialogues.

However, turn detection still faces significant challenges in both methodology and data. Existing approaches struggle to balance accuracy and efficiency, particularly in real-time, noisy, and overlapping speech scenarios. For instance, Ten Turn relies on ASR transcripts, and the additional ASR module introduces latency while performing poorly in noisy environments. Smart Turn uses a simple linear layer for prediction, which makes it less effective in handling complex conversational scenarios. Although Easy Turn generates accurate outputs, it still faces latency issues due to the need to first output ASR results, and it has limited capacity to model complex acoustic information.
Moreover, existing open-source dialogue corpora generally lack fine-grained turn-taking annotations, which limits the ability to reliably model and evaluate turn detection. Although some dialogue datasets~\cite{kraaij2005ami,yang2022open} provide partial turn annotations, these datasets still fall short of meeting the demands of modern dialogue systems—especially in real-world scenarios involving multiple participants, background noise, and natural speech interactions. Furthermore, many turn detection datasets and full-duplex interaction benchmarks~\cite{lin2025full,peng2025fd} are not derived from natural dialogues and rarely include realistic interaction structures with speech overlap, leading to a mismatch between offline benchmarks and real-world deployment.

To address these limitations, we propose \textbf{FastTurn}, a unified framework for low-latency and robust turn detection. FastTurn incorporates a streaming Connectionist Temporal Classification (CTC) module~\cite{graves2006connectionist} to enable fast decoding from partial observations, thereby reducing the latency accumulation commonly introduced by ASR-based cascaded pipelines. By directly integrating acoustic features with learned decision modeling, the framework mitigates the information loss inherent in text-only approaches while preserving real-time responsiveness. In addition, we release the FastTurn test set~\footnote{\url{https://github.com/qualialabsAI/SmoothConv}}, specifically designed to capture authentic turn transitions and overlapping speech. The dataset includes challenging conversational phenomena such as backchannels, pauses, pitch variations, and environmental noise. Based on this evaluation set, we can systematically analyze interaction patterns prone to interruption errors, bridging the gap between controlled offline benchmarks and real-world deployment conditions. Extensive experiments demonstrate that FastTurn achieves consistently higher decision accuracy while substantially reducing interruption latency compared with representative baselines. Notably, the framework maintains robust performance under challenging scenarios involving conversational backchannels and environmental noise, validating its effectiveness for practical full-duplex spoken dialogue systems.

\section{FastTurn}
\begin{figure}[t]
    \centering
    \includegraphics[trim={0 0.4cm 0 0.4cm}, clip, width=\columnwidth]{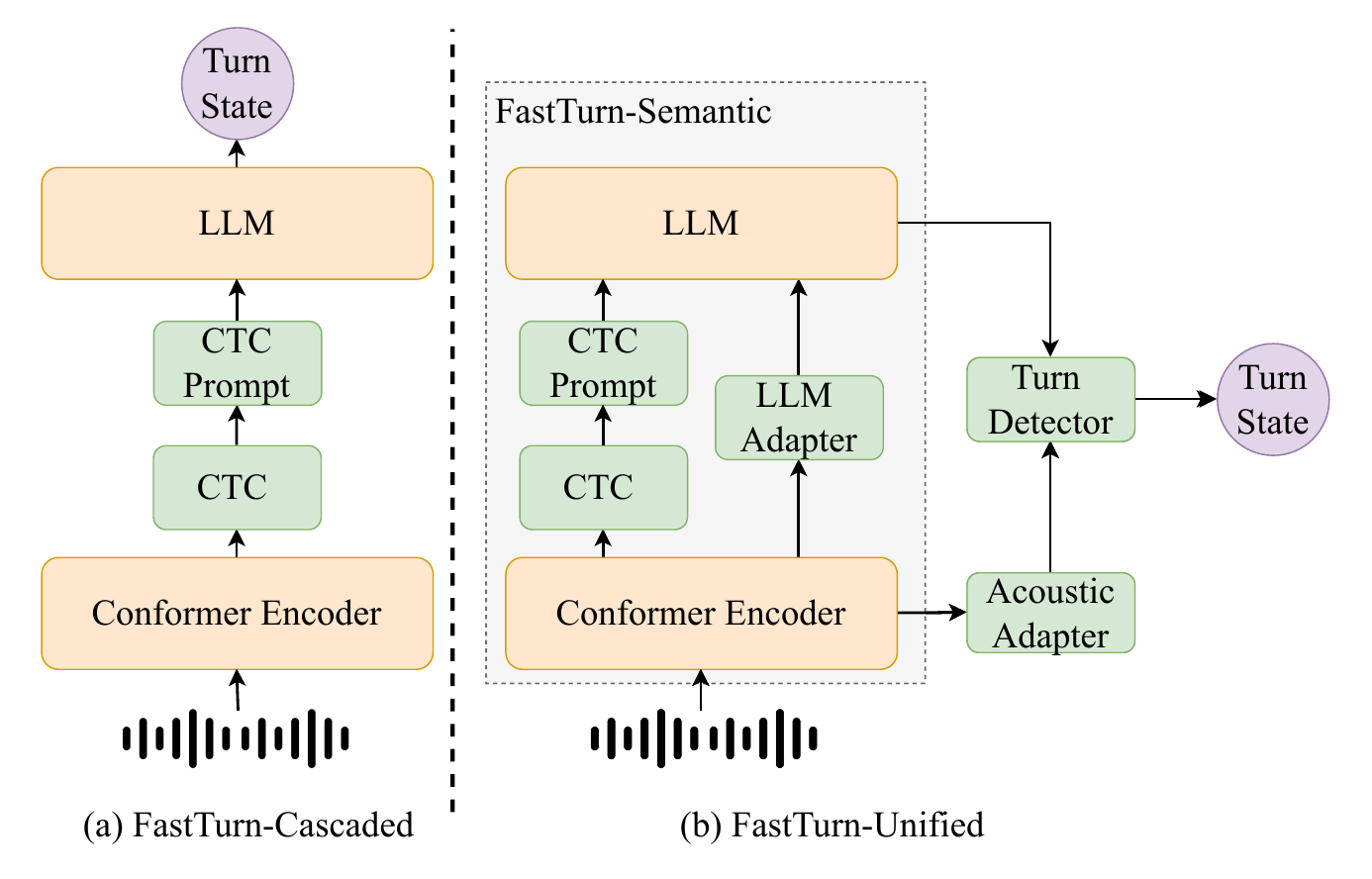}
    \caption{Model architecture}
    \label{fig:fastturn}
\end{figure}
\subsection{Architecture}

As shown in Figure \ref{fig:fastturn}, the framework consists of three components: FastTurn-Semantic, an acoustic adapter, and a turn detector. The architecture is developed in three steps. First, we use FastTurn-Cascaded to route a fast CTC transcript to an LLM for low-latency decisions, then progressively introduce speech-derived cues to enhance robustness.

\textbf{FastTurn-Cascaded}. Turn decisions rely on a transcript, but generating it introduces decoding latency. To minimize this, we introduce a CTC branch for fast alignment and greedy decoding, enabling streaming transcription. The transcription is formatted as a CTC prompt and fed into the LLM (Qwen3-0.6B)~\cite{yang2025qwen3} for turn prediction, providing explicit cues with minimal decoding overhead. However, as the LLM input is dominated by the CTC transcript, predictions are sensitive to CTC errors, especially in the presence of speech overlap and noise.

\textbf{FastTurn-Semantic}. To reduce reliance on transcript quality, we extend the FastTurn-Cascaded design by incorporating speech-derived features into the LLM. The Conformer encoder extracts high-level acoustic representations, which, along with the CTC prompt, are projected into the LLM input space through an LLM adapter, ensuring feature alignment. The LLM then uses both the CTC prompt and aligned acoustic embeddings for turn-related reasoning. This approach allows FastTurn-Semantic to mitigate CTC errors while preserving the latency advantage of early textual conditioning.

\textbf{FastTurn-Unified}. Finally, we enhance robustness by fusing semantic and streaming acoustic cues before the final decision. Intermediate hidden states from the Conformer encoder are processed by an acoustic adapter to extract fine-grained acoustic features. These are fused with the LLM's hidden states and forwarded to the turn detector, implemented as a multi-layer perceptron. The detector predicts whether the current speech segment is a complete turn. By combining streaming acoustic cues from CTC, LLM-conditioned semantic modeling, and acoustic-semantic fusion, the framework improves turn prediction when lexical evidence is ambiguous and prosodic cues are critical, while maintaining efficient inference.
\begin{figure*}[htbp]
    \centering
    \includegraphics[trim={0.5cm 0.45cm 0.55cm 0.1cm}, clip, width=0.95\textwidth]{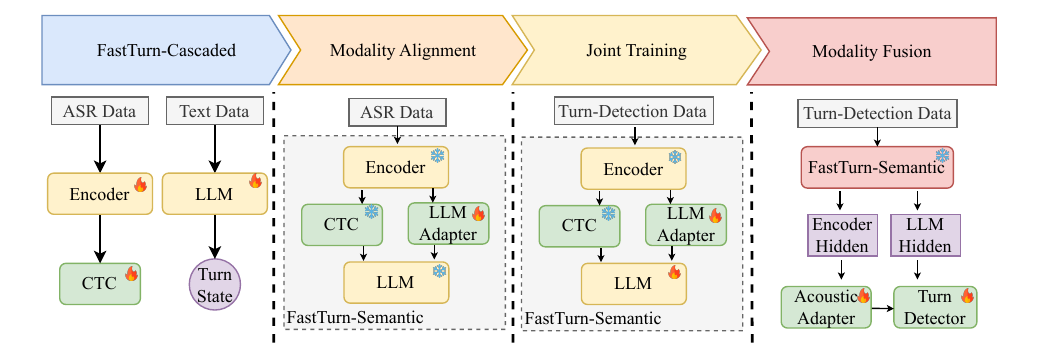}
    \caption{Training Strategy}
    \label{fig:train}
\end{figure*}

\subsection{Train}

Accurate turn detection relies on the understanding of semantic information, while acoustic cues are equally essential in complex scenarios. To fully leverage semantic information and effectively handle challenges such as noise or overlapping speech, we adopt a four-stage training pipeline, as shown in Figure \ref{fig:train}, to stabilize the optimization process and establish speech-text alignment, thereby improving turn prediction accuracy.

\textbf{Semantic Pretraining}. To obtain reliable semantic information, we first train the Conformer encoder and the CTC branch on ASR data. Simultaneously, we fine-tune the LLM on text-only data to better adapt it for the turn detection task. By inserting the turn state as a special token in the input sequence, we reduce the number of tokens generated by the LLM, thereby improving its efficiency for this task.

\textbf{Modality Alignment}.
To address the limitations of CTC in noisy or overlapping speech conditions, we train the LLM adapter under the ASR objective to map the encoder outputs into the LLM input space. This ensures consistency between speech representations and textual supervision in a shared semantic space.

\textbf{Joint Training}. To activate the LLM’s inherent capability for turn detection, we jointly train the LLM and LLM adapter, conditioning the prediction on both acoustic embeddings and the CTC prompt. To prevent overfitting to the CTC branch and preserve the LLM's language modeling ability, we apply prompt dropout: with a probability of $p<0.5$, the CTC prompt is randomly dropped during training, encouraging better generalization.

\textbf{Modality Fusion}. To further improve the robustness and accuracy of turn detection, we fuse semantic and acoustic cues. We train the acoustic adapter and turn detector on the same dataset, combining Conformer representations with LLM hidden states. This fusion enables joint modeling of prosodic and semantic cues, significantly enhancing the accuracy of turn boundary prediction and increasing the overall robustness of the system.

\subsection{FastTurn test set}

Current open-source dialogue corpora often lack detailed turn-taking annotations, limiting the development of reliable turn detection models. To address this issue, we collected high-quality dual-channel real human-to-human dialogue data and, after precise annotation, constructed our test set. The annotations include detailed labels such as speaker identities, emotions, timestamps, turn boundaries, paralinguistic cues (such as pauses, overlaps, and backchannels), and transcriptions. These annotations provide a comprehensive understanding of interaction structure, temporal alignment, and interruption behaviors. By combining dual-channel audio with these rich, multi-dimensional annotations and transcriptions, the test set serves as an important resource for research on dialogue coordination, interruption modeling, and full-duplex systems, aimed at accurately capturing turn transitions and interaction flow in natural dialogues.

\begin{table}[!ht]
  \caption{Statistics of the FastTurn test set.}
  \label{tab:testset_statistics}
  \centering
  \begin{tabular}{l l r r}
    \toprule
    \textbf{Turn State} & \textbf{Source} & \textbf{Samples} & \textbf{Duration (h)} \\
    \midrule
    Complete    & real-world  & 14709 & 9.64 \\
    Incomplete  & real-world  & 3643  & 2.15 \\
    Backchannel & real-world  & 3080  & 0.42 \\
    Wait        & Synthesized  & 1000  & 0.71 \\
    \bottomrule
  \end{tabular}
\end{table}

To evaluate turn-state prediction, we construct an evaluation set consisting of segments from real-world data and 1,000 synthetically generated \textit{wait} state samples, as shown in Table~\ref{tab:testset_statistics}. Since the \textit{wait} state is rare in natural conversations, we supplement the set with 1,000 samples generated using DeepSeek V3~\cite{liu2024deepseek} for text and IndexTTS2~\cite{zhou2025indextts2} for audio synthesis.

\section{Experiments}
\subsection{Datasets}

\textbf{ASR Task}. We use large-scale open-source corpora and internal datasets, including AISHELL-1~\cite{bu2017aishell}, AISHELL-2~\cite{du2018aishell}, WenetSpeech~\cite{zhang2022wenetspeech}, LibriSpeech~\cite{panayotov2015librispeech}, GigaSpeech~\cite{chen2021gigaspeech}, and MLS~\cite{pratap2020mls}, totaling over 30,000 hours of Chinese and English speech to support robust feature learning.

\noindent \textbf{Turn Detection Task.} We use the Easy Turn training set, augmented with internal conversational data and synthetic corpora. Dialogue texts are generated by Qwen3-32B~\cite{yang2025qwen3} and DeepSeek-v3~\cite{liu2024deepseek}, then synthesized into speech using Indextts2~\cite{zhou2025indextts2}. To generate \textit{complete} and \textit{incomplete} states, we use forced alignment to extract word-level timestamps and create negative samples by truncating complete turns at random temporal positions, ensuring linguistic incompleteness through filtering.

\subsection{Experimental setup}
The model architecture consists of a 12-layer Conformer encoder~\cite{gulati2020conformer} with approximately 80 million parameters, using 8 attention heads and a convolutional module with a kernel size of 8. Both the LLM and acoustic adapters are based on 4-layer Transformer architectures, each with approximately 24 million parameters. The turn detector is a 3-layer multi-layer perceptron (MLP) that predicts turn boundaries using fused features from the Conformer encoder and LLM.
All experiments are conducted on 8 NVIDIA A6000 GPUs. For the ASR task, the learning rate is set to $1 \times 10^{-4}$ with a warm-up period of 8,000 steps, and training lasts for 80,000 steps. For turn detection, LLM fine-tuning is done with a learning rate of $1 \times 10^{-5}$ for 2 epochs, followed by joint training and modality fusion with learning rates of $5 \times 10^{-6}$ and $1 \times 10^{-4}$, respectively, each trained for 11,000 steps.
\subsection{Evaluation metrics}
To evaluate model performance in full-duplex conversational scenarios, we employ three primary metrics: Accuracy, Miss Rate, and False Alarm Rate. These are derived from turn-state classification results, where True Positives (TP) and True Negatives (TN) denote correct predictions, and False Positives (FP) and False Negatives (FN) denote errors.
\begin{equation}
\text{Accuracy} = \frac{\text{TP} + \text{TN}}{\text{TP} + \text{TN} + \text{FP} + \text{FN}}.
\end{equation}

\begin{equation}
\text{Miss Rate} = \frac{\text{FN}}{\text{TP} + \text{FN}}.
\end{equation}

\begin{equation}
\text{False Alarm Rate} = \frac{\text{FP}}{\text{FP} + \text{TN}}.
\end{equation}

\subsection{Main results}

\begin{table*}[t]
\caption{Turn-detection performance on the FastTurn test set. Higher Accuracy ($\uparrow$) and lower Miss Rate ($\downarrow$) and False Alarm Rate ($\downarrow$) are better. Para. denotes Paraformer, \colorbox{lightgreen}{green} rows indicate our models. \textbf{Bold} numbers show best results; \underline{underlined} numbers show second-best results.}
  \label{tab:fastturn_results}
  \centering
  \setlength{\tabcolsep}{4pt}
  \resizebox{0.98\linewidth}{!}{
  \begin{tabular}{l|ccc|ccc|ccc|ccc}
    \toprule
    \multirow{2}{*}{Model} 
    & \multicolumn{3}{c|}{Complete} 
    & \multicolumn{3}{c|}{Incomplete} 
    & \multicolumn{3}{c|}{Backchannel} 
    & \multicolumn{3}{c}{Wait} \\
    \cmidrule(lr){2-4} \cmidrule(lr){5-7} \cmidrule(lr){8-10} \cmidrule(lr){11-13}
    & Acc $\uparrow$ & Miss $\downarrow$ & FA $\downarrow$
    & Acc $\uparrow$ & Miss $\downarrow$ & FA $\downarrow$
    & Acc $\uparrow$ & Miss $\downarrow$ & FA $\downarrow$
    & Acc $\uparrow$ & Miss $\downarrow$ & FA $\downarrow$ \\
    \midrule

    Para.~\cite{gao2022paraformer}+Ten Turn~\footnote{\url{https://github.com/TEN-framework/ten-turn-detection}} & 71.52 & \underline{28.71} & 28.34 & 58.27 & 32.70 & 47.15 & -- & -- & -- & 98.15 & 2.78 & 0.31 \\
    Smart Turn~\footnote{\url{https://github.com/pipecat-ai/Smart-Turn}}          & 49.21 & 49.97 & 51.93 & 49.21 & 51.93 & 49.97 & -- & -- & -- & -- & -- & -- \\
    
Easy Turn~\cite{DBLP:journals/corr/abs-2509-23938}          & \underline{80.10} & \underline{21.93} & 15.46 & \textbf{82.28} & 35.21 & \textbf{14.14} & \underline{93.91} & \textbf{6.40} & 6.03 & \underline{98.64} & \textbf{2.20} & \textbf{0.05} \\

    \rowcolor{lightgreen}
    FastTurn-Cascaded           & 73.26 & 34.60 & \textbf{12.29} & 65.95 & \textbf{24.11} & 36.49 & 86.62 & 66.24 & \textbf{3.56} & 97.21 & 4.89 & 0.21 \\
    \rowcolor{lightgreen}
    \ \ + FastTurn-Semantic           & 79.69 & 22.67 & 15.17 & 76.41 & \underline{32.03} & 21.87 & 89.55 & 43.73 & \underline{4.87} & 98.57 & 3.79 & \underline{0.18} \\
    \rowcolor{lightgreen}
    \ \ \ \ + FastTurn-Unified           & \textbf{81.64} & \textbf{14.53} & \underline{14.92} & \underline{81.01} & 35.71 & \underline{15.57} & \textbf{93.93} & \underline{7.68} & 5.63 & \textbf{98.75} & \underline{2.31} & 0.39 \\
    
    \bottomrule
  \end{tabular}}
\end{table*}

\begin{table*}[htb]
\centering
\caption{Turn-detection performance comparison across test sets. Acc denotes overall accuracy (\%), Lat. denotes average latency (ms), and Params denotes the number of parameters (M). 
``Com'' and ``Inc'' denote \textit{complete} and \textit{incomplete} detection, respectively.}
\label{tab:turn_detection_smart_comparison}
\setlength{\tabcolsep}{4pt}
\begin{tabular}{l c | cc | cc | cc | ccc | ccc}
    \toprule
    Model & Params 
    & \multicolumn{2}{c|}{Smart Turn (zh)} 
    & \multicolumn{2}{c|}{Easy Turn} 
    & \multicolumn{2}{c}{FastTurn } 
    & \multicolumn{3}{c|}{Smart Turn (en)-Com} 
    & \multicolumn{3}{c}{Smart Turn (en)-Inc} \\
    \cmidrule(lr){3-4}\cmidrule(lr){5-6}\cmidrule(lr){7-8}\cmidrule(lr){9-11}\cmidrule(lr){12-14}
     & 
     & Acc & Lat. 
     & Acc & Lat. 
     & Acc & Lat. 
     & Acc $\uparrow$ & Miss $\downarrow$ & FA $\downarrow$
     & Acc $\uparrow$ & Miss $\downarrow$ & FA $\downarrow$ \\
    \midrule
    Para.+Ten Turn & 7220 
    & \underline{83.10} & \underline{124.3}
    & 86.00 & 212.0 
    & 51.97 & \textbf{114.8}
    & \underline{79.07} & 15.27 & \underline{26.36} & \textbf{79.13} & \underline{26.44} & 15.06 \\
    Smart Turn & 32 
    & \textbf{90.53} & \textbf{70.22}
    & 76.86 & \textbf{62.28}
    & 49.21 & \textbf{116.9} 
    & \textbf{94.71} & \textbf{4.72} & \textbf{5.84} & \textbf{94.71} & \textbf{5.84} & \textbf{4.72} \\
    Easy Turn & 850 
    & 57.16 & 687.8
    & \textbf{96.38} & 355.9 
    & \underline{78.05} & 297.1 
    & -- & -- & -- & -- & -- & -- \\
    \rowcolor{lightgreen}
    FastTurn-Cascaded & 650
    & 75.42 & 150.1 
    & \underline{96.13} & 153.1 
    & 62.50 & 126.3 
    & 76.18 & \underline{4.93} & 41.99 & 76.09 & 42.76 & \underline{4.30} \\
    \rowcolor{lightgreen}
    \ \ +FastTurn-Unified & 700 
    & 76.58 & 139.0 
    & 94.50 & \underline{136.4} 
    & \textbf{79.62} & 120.1 
    & 77.34 & 5.24 & 39.40 & \underline{77.35} & 40.01 & 4.59 \\
    \bottomrule
\end{tabular}
\end{table*}

To evaluate the model's performance in complex conversational scenarios, Table~\ref{tab:fastturn_results} presents the results of the FastTurn model and baseline models. The results indicate that models relying solely on semantic information, such as Paraformer+Ten Turn and FastTurn-Cascaded, perform poorly in both the \textit{Complete} and \textit{Incomplete} categories, exhibiting low accuracy along with high miss and false alarm rates. In contrast, models that better integrate semantic and acoustic information, such as Easy Turn and FastTurn-Unified, perform better. Notably, FastTurn-Unified achieves the best performance across all categories, highlighting the importance of deeply combining semantic and acoustic features to improve model performance.

We evaluate the model's accuracy and latency across multiple test sets to assess its performance and real-time processing capabilities. Table~\ref{tab:turn_detection_smart_comparison} shows results for the Smart Turn, Easy Turn, and FastTurn test sets. The Smart Turn model outperforms others on its test set due to mismatches in data and label categories. The test set, with only two categories (\textit{\textquotedblleft complete\textquotedblright} and \textit{\textquotedblleft incomplete\textquotedblright}), does not fully align with our multi-class model, and its simplified data contributes to its better performance.
The Easy Turn test set, with 800 samples and no background noise, relies on semantic cues, leading to strong performance from FastTurn-Cascaded. However, FastTurn performs slightly worse due to a small sample size and sensitivity to variations. The FastTurn test set, with more echo signals and acoustic ambiguity, presents a greater challenge for turn-state modeling.
In terms of latency, Smart Turn's simplified design results in low latency but poorer performance in complex scenarios. FastTurn-Unified achieves lower latency than both Easy Turn and FastTurn-Cascaded while maintaining similar or better accuracy. For the English subset, results met expectations, but did not surpass Paraformer+Ten Turn\cite{gao2022paraformer}. Limited optimization and English dialogue data affected the model's performance.

\subsection{ASR results}
\begin{table}[htb]
\centering
\caption{Recognition performance on evaluation sets: Chinese CER (\%) and English WER (\%). For LLM decoding, variants differ only in the LLM adapter architecture.}
\label{tab:asr_results}
\setlength{\tabcolsep}{3pt}  
\renewcommand{\arraystretch}{1.2}  
\resizebox{1.0\linewidth}{!}{
\begin{tabular}{l l ccc}  
\hline
Decoding & LLM Adapter & LibriClean & TestNet & AISHELL-1 \\
\hline
CTC greedy & -- & \underline{7.06} & \textbf{9.52} & \textbf{2.33} \\
\hline
LLM & 2L MLP & 14.09 & 16.80 & 6.45 \\
LLM & 2L Transformer & 7.19 & \underline{10.74} & 5.31 \\
LLM & 4L Transformer & \textbf{5.56} & \underline{10.74} & \underline{3.69} \\
\hline
\end{tabular}}
\end{table}
As shown in Table~\ref{tab:asr_results}, LLM-based autoregressive decoding achieves performance close to CTC decoding on most evaluation sets. During training, only the adapter parameters are updated while all other components remain frozen. This constrained setting limits full adaptation to the ASR objective and results in a consistent yet moderate gap compared with CTC decoding. Nevertheless, the aligned representations effectively inject high-level semantic information into the LLM, enabling competitive autoregressive decoding despite restricted parameter updates.

\subsection{Ablation study}

As shown in Table~\ref{tab:fastturn_results}, FastTurn-Semantic improves turn detection performance over FastTurn-Cascaded by reducing reliance on transcript quality and incorporating speech-derived features, which helps compensate for CTC errors in noisy or overlapping speech conditions. FastTurn-Unified further demonstrates the effectiveness of combining semantic and acoustic cues for real-time turn detection.

\section{Conclusion}

This paper presents FastTurn, a framework for efficient turn detection in full-duplex systems. By utilizing fast CTC decoding and integrating acoustic features, FastTurn reduces latency and enhances robustness. We release a comprehensive test set to promote research on conversational turn-taking and speech overlap phenomena, specifically designed to capture realistic interaction dynamics. FastTurn effectively handles complex conversational patterns, such as echo signals and speech overlap, while maintaining high accuracy and low latency. Experimental results demonstrate that FastTurn exhibits strong robustness under challenging acoustic conditions, making it a promising solution for real-time, scalable turn detection. Future work will focus on optimizing the model's performance and extending its application to more dynamic conversational scenarios.

\bibliographystyle{IEEEtran}
\bibliography{mybib}

@article{gao2022paraformer,
  title={Paraformer: Fast and accurate parallel transformer for non-autoregressive end-to-end speech recognition},
  author={Gao, Zhifu and Zhang, Shiliang and McLoughlin, Ian and Yan, Zhijie},
  journal={arXiv preprint arXiv:2206.08317},
  year={2022}
}

@article{geng2025osum,
  title={Osum-echat: Enhancing end-to-end empathetic spoken chatbot via understanding-driven spoken dialogue},
  author={Geng, Xuelong and Shao, Qijie and Xue, Hongfei and Wang, Shuiyuan and Xie, Hanke and Guo, Zhao and Zhao, Yi and Li, Guojian and Tian, Wenjie and Wang, Chengyou and others},
  journal={arXiv preprint arXiv:2508.09600},
  year={2025}
}

@article{yang2025qwen3,
  title={Qwen3 technical report},
  author={Yang, An and Li, Anfeng and Yang, Baosong and Zhang, Beichen and Hui, Binyuan and Zheng, Bo and Yu, Bowen and Gao, Chang and Huang, Chengen and Lv, Chenxu and others},
  journal={arXiv preprint arXiv:2505.09388},
  year={2025}
}

@article{liu2024deepseek,
  title={Deepseek-v3 technical report},
  author={Liu, Aixin and Feng, Bei and Xue, Bing and Wang, Bingxuan and Wu, Bochao and Lu, Chengda and Zhao, Chenggang and Deng, Chengqi and Zhang, Chenyu and Ruan, Chong and others},
  journal={arXiv preprint arXiv:2412.19437},
  year={2024}
}

@article{zhou2025indextts2,
  title={Indextts2: A breakthrough in emotionally expressive and duration-controlled auto-regressive zero-shot text-to-speech},
  author={Zhou, Siyi and Zhou, Yiquan and He, Yi and Zhou, Xun and Wang, Jinchao and Deng, Wei and Shu, Jingchen},
  journal={arXiv preprint arXiv:2506.21619},
  year={2025}
}

@article{defossez2024moshi,
  title={Moshi: a speech-text foundation model for real-time dialogue},
  author={D{\'e}fossez, Alexandre and Mazar{\'e}, Laurent and Orsini, Manu and Royer, Am{\'e}lie and P{\'e}rez, Patrick and J{\'e}gou, Herv{\'e} and Grave, Edouard and Zeghidour, Neil},
  journal={arXiv preprint arXiv:2410.00037},
  year={2024}
}

@article{yu2024salmonn,
  title={Salmonn-omni: A codec-free llm for full-duplex speech understanding and generation},
  author={Yu, Wenyi and Wang, Siyin and Yang, Xiaoyu and Chen, Xianzhao and Tian, Xiaohai and Zhang, Jun and Sun, Guangzhi and Lu, Lu and Wang, Yuxuan and Zhang, Chao},
  journal={arXiv preprint arXiv:2411.18138},
  year={2024}
}

@article{wang2024freeze,
  title={Freeze-omni: A smart and low latency speech-to-speech dialogue model with frozen llm},
  author={Wang, Xiong and Li, Yangze and Fu, Chaoyou and Shen, Yunhang and Xie, Lei and Li, Ke and Sun, Xing and Ma, Long},
  journal={arXiv preprint arXiv:2411.00774},
  year={2024}
}

@article{zeng2024glm,
  title={Glm-4-voice: Towards intelligent and human-like end-to-end spoken chatbot},
  author={Zeng, Aohan and Du, Zhengxiao and Liu, Mingdao and Wang, Kedong and Jiang, Shengmin and Zhao, Lei and Dong, Yuxiao and Tang, Jie},
  journal={arXiv preprint arXiv:2412.02612},
  year={2024}
}

@article{lin2025full,
  title={Full-duplex-bench: A benchmark to evaluate full-duplex spoken dialogue models on turn-taking capabilities},
  author={Lin, Guan-Ting and Lian, Jiachen and Li, Tingle and Wang, Qirui and Anumanchipalli, Gopala and Liu, Alexander H and Lee, Hung-yi},
  journal={arXiv preprint arXiv:2503.04721},
  year={2025}
}

@article{peng2025fd,
  title={FD-Bench: A Full-Duplex Benchmarking Pipeline Designed for Full Duplex Spoken Dialogue Systems},
  author={Peng, Yizhou and Chao, Yi-Wen and Ng, Dianwen and Ma, Yukun and Ni, Chongjia and Ma, Bin and Chng, Eng Siong},
  journal={arXiv preprint arXiv:2507.19040},
  year={2025}
}

@inproceedings{zhang2022wenetspeech,
  title={Wenetspeech: A 10000+ hours multi-domain mandarin corpus for speech recognition},
  author={Zhang, Binbin and Lv, Hang and Guo, Pengcheng and Shao, Qijie and Yang, Chao and Xie, Lei and Xu, Xin and Bu, Hui and Chen, Xiaoyu and Zeng, Chenchen and others},
  booktitle={IEEE International Conference on Acoustics, Speech and Signal Processing (ICASSP)},
  pages={6182--6186},
  year={2022},
  organization={IEEE}
}

@inproceedings{bu2017aishell,
  title={Aishell-1: An open-source mandarin speech corpus and a speech recognition baseline},
  author={Bu, Hui and Du, Jiayu and Na, Xingyu and Wu, Bengu and Zheng, Hao},
  booktitle={2017 20th conference of the oriental chapter of the international coordinating committee on speech databases and speech I/O systems and assessment (O-COCOSDA)},
  pages={1--5},
  year={2017},
  organization={IEEE}
}

@article{du2018aishell,
  title={Aishell-2: Transforming mandarin asr research into industrial scale},
  author={Du, Jiayu and Na, Xingyu and Liu, Xuechen and Bu, Hui},
  journal={arXiv preprint arXiv:1808.10583},
  year={2018}
}

@inproceedings{panayotov2015librispeech,
  title={Librispeech: an asr corpus based on public domain audio books},
  author={Panayotov, Vassil and Chen, Guoguo and Povey, Daniel and Khudanpur, Sanjeev},
  booktitle={2015 IEEE international conference on acoustics, speech and signal processing (ICASSP)},
  pages={5206--5210},
  year={2015},
  organization={IEEE}
}

@inproceedings{graves2006connectionist,
  title={Connectionist temporal classification: labelling unsegmented sequence data with recurrent neural networks},
  author={Graves, Alex and Fern{\'a}ndez, Santiago and Gomez, Faustino and Schmidhuber, J{\"u}rgen},
  booktitle={Proceedings of the 23rd international conference on Machine learning},
  pages={369--376},
  year={2006}
}

@article{fu2025vita,
  title={Vita-1.5: Towards gpt-4o level real-time vision and speech interaction},
  author={Fu, Chaoyou and Lin, Haojia and Wang, Xiong and Zhang, Yi-Fan and Shen, Yunhang and Liu, Xiaoyu and Cao, Haoyu and Long, Zuwei and Gao, Heting and Li, Ke and others},
  journal={arXiv preprint arXiv:2501.01957},
  year={2025}
}

@article{chen2025fireredchat,
  title={Fireredchat: A pluggable, full-duplex voice interaction system with cascaded and semi-cascaded implementations},
  author={Chen, Junjie and Hu, Yao and Li, Junjie and Li, Kangyue and Liu, Kun and Li, Wenpeng and Li, Xu and Li, Ziyuan and Shen, Feiyu and Tang, Xu and others},
  journal={arXiv preprint arXiv:2509.06502},
  year={2025}
}

@article{chu2024qwen2,
  title={Qwen2-audio technical report},
  author={Chu, Yunfei and Xu, Jin and Yang, Qian and Wei, Haojie and Wei, Xipin and Guo, Zhifang and Leng, Yichong and Lv, Yuanjun and He, Jinzheng and Lin, Junyang and others},
  journal={arXiv preprint arXiv:2407.10759},
  year={2024}
}

@article{yang2022open,
  title={Open source magicdata-ramc: A rich annotated mandarin conversational (ramc) speech dataset},
  author={Yang, Zehui and Chen, Yifan and Luo, Lei and Yang, Runyan and Ye, Lingxuan and Cheng, Gaofeng and Xu, Ji and Jin, Yaohui and Zhang, Qingqing and Zhang, Pengyuan and others},
  journal={arXiv preprint arXiv:2203.16844},
  year={2022}
}

@inproceedings{liu2020towards,
  title={Towards building an intelligent chatbot for customer service: Learning to respond at the appropriate time},
  author={Liu, Che and Jiang, Junfeng and Xiong, Chao and Yang, Yi and Ye, Jieping},
  booktitle={Proceedings of the 26th ACM SIGKDD international conference on Knowledge Discovery \& Data Mining},
  pages={3377--3385},
  year={2020}
}

@inproceedings{lu2025duplexmamba,
  title={Duplexmamba: Enhancing real-time speech conversations with duplex and streaming capabilities},
  author={Lu, Xiangyu and Xu, Wang and Wang, Haoyu and Zhou, Hongyun and Zhao, Haiyan and Zhu, Conghui and Zhao, Tiejun and Yang, Muyun},
  booktitle={CCF International Conference on Natural Language Processing and Chinese Computing},
  pages={62--74},
  year={2025},
  organization={Springer}
}

@article{xie2024mini,
  title={Mini-omni: Language models can hear, talk while thinking in streaming},
  author={Xie, Zhifei and Wu, Changqiao},
  journal={arXiv preprint arXiv:2408.16725},
  year={2024}
}

@article{leviathan2018google,
  title={Google Duplex: An AI system for accomplishing real-world tasks over the phone},
  author={Leviathan, Yaniv and Matias, Yossi},
  journal={Google AI blog},
  volume={8},
  year={2018}
}

@article{chen2021gigaspeech,
  title={Gigaspeech: An evolving, multi-domain asr corpus with 10,000 hours of transcribed audio},
  author={Chen, Guoguo and Chai, Shuzhou and Wang, Guanbo and Du, Jiayu and Zhang, Wei-Qiang and Weng, Chao and Su, Dan and Povey, Daniel and Trmal, Jan and Zhang, Junbo and others},
  journal={arXiv preprint arXiv:2106.06909},
  year={2021}
}

@article{pratap2020mls,
  title={Mls: A large-scale multilingual dataset for speech research},
  author={Pratap, Vineel and Xu, Qiantong and Sriram, Anuroop and Synnaeve, Gabriel and Collobert, Ronan},
  journal={arXiv preprint arXiv:2012.03411},
  year={2020}
}

@inproceedings{huang2024audiogpt,
  title={Audiogpt: Understanding and generating speech, music, sound, and talking head},
  author={Huang, Rongjie and Li, Mingze and Yang, Dongchao and Shi, Jiatong and Chang, Xuankai and Ye, Zhenhui and Wu, Yuning and Hong, Zhiqing and Huang, Jiawei and Liu, Jinglin and others},
  booktitle={Proceedings of the AAAI Conference on Artificial Intelligence},
  volume={38},
  number={21},
  pages={23802--23804},
  year={2024}
}

@article{gulati2020conformer,
  title={Conformer: Convolution-augmented transformer for speech recognition},
  author={Gulati, Anmol and Qin, James and Chiu, Chung-Cheng and Parmar, Niki and Zhang, Yu and Yu, Jiahui and Han, Wei and Wang, Shibo and Zhang, Zhengdong and Wu, Yonghui and others},
  journal={arXiv preprint arXiv:2005.08100},
  year={2020}
}

@inproceedings{kraaij2005ami,
  title={The AMI meeting corpus},
  author={Kraaij, Wessel and Hain, Thomas and Lincoln, Mike and Post, Wilfried},
  booktitle={Proc. International Conference on Methods and Techniques in Behavioral Research},
  pages={1--4},
  year={2005}
}

@inproceedings{DBLP:journals/corr/abs-2509-23938,
  author       = {Guojian Li and
                  Chengyou Wang and
                  Hongfei Xue and
                  Shuiyuan Wang and
                  Dehui Gao and
                  Zihan Zhang and
                  Yuke Lin and
                  Wenjie Li and
                  Longshuai Xiao and
                  Zhonghua Fu and
                  Lei Xie},
  title        = {Easy Turn: Integrating Acoustic and Linguistic Modalities for Robust
                  Turn-Taking in Full-Duplex Spoken Dialogue Systems},
  booktitle={IEEE International Conference on Acoustics, Speech and Signal Processing (ICASSP)},
  year={2026},
}

\end{document}